\documentclass{aa}
\def\kato{\rule[-1.25ex]{0cm}{1.25ex}}
\def\pano{\rule[0.0ex]{0cm}{2.5ex}}

\def\h2{H{\small\,II}}

\newcommand{\sbu}{mag arcsec$^{-2}$}

\begin{document}

  \thesaurus{  04          
              (11.01.1;    
               11.05.2;    
               11.06.1;    
               11.09.4;    
               11.16.1;    
               11.19.3;    
               11.19.5; )} 

   \title{Is I Zw 18 a young galaxy?}

   \author{Y. I. Izotov
    \inst{1}
    \and
    P. Papaderos \inst{2}
    \and
    T. X. Thuan \inst{3}
    \and
    K. J. Fricke \inst{2}
    \and
    C. B. Foltz \inst{4}
    \and
    N. G. Guseva \inst{1}}
   \offprints{Y. Izotov}
   \institute{    Main Astronomical Observatory
                 of National Academy of Sciences of Ukraine,
                 Goloseevo, 252650 Kiev-22, Ukraine
              \and
                 Universit\"ats--Sternwarte, Geismarlandstrasse 11,
                 D--37083 G\"ottingen, Germany
              \and
                 Astronomy Department, 
                 University of Virginia, 
                 Charlottesville, VA 22903, USA
              \and
                 Multiple Mirror Telescope Observatory, 
                 University of Arizona, 
                 Tucson, AZ 85721, USA}
\date{Received ; accepted }
\maketitle

\markboth {Y. Izotov et al.: Is I Zw 18 a young galaxy?}{}

\begin{abstract}
Hubble Space Telescope (HST)\footnote[1]{Based on observations
obtained with the NASA/ESA {\sl Hubble Space Telescope} through the Space
Telescope Science Institute, which is operated by AURA, Inc. under NASA 
contract NAS5-26555.} colour - magnitude diagrams in $B$, $V$ and $R$
along with long-slit Multiple 
Mirror Telescope (MMT)\footnote[2]{Ground-based spectroscopic observations 
presented herein were obtained with the Multiple Mirror Telescope, a 
facility operated jointly by the Smithsonian Institution and the 
University of Arizona.} spectrophotometric data are used to investigate the  
evolutionary status of the nearby blue compact dwarf (BCD) galaxy I\ Zw\ 18.
We find that the distance to I Zw 18 should be as high as 20 Mpc,
twice the previously accepted distance, to be consistent with
existing observational data on the galaxy: colour-magnitude diagrams, 
the high ionization state of the gas and presence of WR stars in the main body, 
and the ionization state of the C component. The 
spectral energy distribution (SED) of the main 
body of I\ Zw\ 18 is consistent with that of a stellar population 
with age $\la$ 5 Myr. However, the presence of large-scale shells observed
around the main body suggests that star formation began $\sim$ 20 Myr at the
NW end and propagated in the SE direction.
Our analysis of colour-magnitude diagrams and of the
spectral energy distribution of the C component implies that star 
formation in this component started $\la$ 100 Myr ago at the NW end, 
propagated to the SE and stopped $\sim$ 15 Myr ago. 
Thus, I Zw 18 is likely to be one of the youngest nearby extragalactic objects.


\end{abstract}

\section{Introduction}
%
I Zw 18 remains the most metal-poor blue compact dwarf (BCD) galaxy known
since its discovery by Sargent \& Searle (1970).
Later spectroscopic observations by Searle \& Sargent (1972), Lequeux et al. 
(1979), French (1980), Kinman \& Davidson
(1981), Pagel et al. (1992), Skillman \& Kennicutt (1993), Martin (1996),
Izotov, Thuan \& Lipovetsky (1997c), Izotov \& Thuan (1998),
V\'ilchez \& Iglesias-P\'aramo (1998), Izotov \& Thuan (1999) and Izotov et al.
(1999) have confirmed its low metallicity with an oxygen abundance of only 
$\sim$ 1/50 the solar value.

\begin{figure}
\vspace{9.cm}
\includegraphics{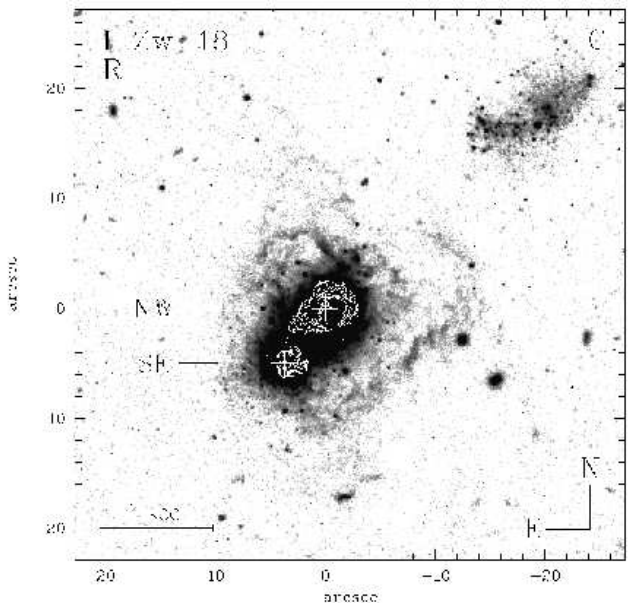}
\caption[]{\label{Fig1}
Hubble Space Telescope WFPC2 exposure of I Zw 18 in the $R$ band. 
The overlayed contours delineate the central H$\alpha$ shells 
surrounding the NW and SE star-forming knots. }

\end{figure}

Zwicky (1966) described I Zw 18 as a double system of compact galaxies,
which are in fact two bright knots of star formation with an angular 
separation of 5\farcs8. 
These two star-forming regions are referred to as the brighter northwest (NW) 
and fainter southeast (SE) components (Fig. \ref{Fig1}) and 
form what we will refer to as the main body.
Later studies by Davidson, Kinman \& Friedman (1989) and Dufour \&
Hester (1990) have revealed a more complex optical morphology.
The most prominent diffuse feature, hereafter component C (Fig. \ref{Fig1}), 
is a blue irregular star-forming region $\sim$ 22\arcsec\ northwest of the NW 
component. Dufour, Esteban \& Casta\~neda (1996a),
van Zee et al. (1998) and Izotov \& Thuan (1998) have shown the C component
to have a systemic radial velocity equal to that of the ionized gas in the 
NW and SE components, thus establishing its physical association to I\ Zw\ 18. 
Furthermore, van Zee et al. (1998) have shown that this component is
embedded in a common H I envelope with the main body. 

Searle \& Sargent (1972) and Hunter \& Thronson (1995) 
have suggested that I\ Zw\ 18 may be a young galaxy,
recently undergoing its first burst of star formation. 
The latter authors concluded from HST images that the colours of 
the diffuse unresolved component surrounding the SE and NW  regions are 
consistent with a population of B and early A stars, i.e. 
with no evidence for older stars. 

Ongoing star formation in the main body of I\ Zw\ 18 is 
implied by the discovery of a population of Wolf-Rayet stars in the NW 
component (Izotov et al. 1997a; Legrand et al. 1997; de Mello et al. 1998). 
Flux-calibrated optical spectra of the C component (Izotov \& Thuan 1998; van 
Zee et al. 1998) reveal a blue continuum with weak Balmer 
absorption features and faint H$\alpha$ and H$\beta$ in emission.
Such spectral features imply that the \h2\ region is ionized by 
a population of early B stars and suggest that the C component 
is older than the main body of I\ Zw\ 18.
Izotov \& Thuan (1998) have suggested an age sequence from the C component 
( $\sim$ 200 Myr ) to the SE region ( $\sim$ 5 Myr ) of active star formation.

Dufour et al. (1996b) have discussed new HST imagery of I\ Zw\ 18, 
including the C component, which is resolved into stars. Based on the
analyses of colour-magnitude diagrams, they concluded that star formation 
in the main body began at least 
30 -- 50 Myr ago and is maintained to the present, as is apparent in the SE 
component. Martin (1996) and Dufour et al. (1996b) have discussed the
properties of expanding superbubbles of ionized gas driven by supernova
explosions and have inferred dynamical ages of respectively 15 -- 27 Myr and 
13 -- 15 Myr. As for the age of the C component, Dufour et al. (1996b) 
found in a $(B-V)$ vs. $V$ colour-magnitude analysis a well defined upper 
stellar main sequence indicating an age of the blue stars of $\sim$ 40 Myr. 
However, numerous faint red stars were also present in the colour-magnitude 
diagram implying an age of 100 -- 300 Myr. 
Dufour et al. (1996b) therefore concluded that the C component
consists of an older stellar population with an age of several hundred Myr, 
but which has experienced recently a modest starburst in its southeastern half
as evidenced by the presence of blue stars and H$\alpha$ emission. 

    Recently, Aloisi, Tosi \& Greggio (1999) have discussed the star formation
history in I Zw 18 using the same HST WFPC2 archival data
( i.e. those by Hunter \& Thronson (1995) and Dufour et al. (1996b)). 
They compared observed 
colour-magnitude diagrams and luminosity functions with synthetic ones and
concluded that there were two episodes of star formation in the main body,
a first episode occuring over the last 0.5 -- 1 Gyr, an age more than 10 
times larger than that derived by Dufour et al. (1996b), and a second episode 
with more intense activity taking place between 15 and 20 Myr ago. 
No star formation has occurred within the last 15 Myr. For the C component, 
Aloisi et al. (1999) estimated an age not exceeding 0.2 Gyr. 

Garnett et al. (1997) have presented measurements of the gas-phase 
C/O abundance ratio in both NW and SE components, based 
on ultraviolet spectroscopy with the Faint Object Spectrograph (FOS)
onboard HST. 
They determined values of log C/O = --0.63 $\pm$ 0.10 for the NW component and 
log C/O = --0.56 $\pm$ 0.09 for the SE component.
These ratios, being significantly higher than in other metal-poor irregular 
galaxies, apparently require that carbon in I\ Zw\ 18 has been 
enriched by an older 
generation of stars. Garnett et al. (1997) concluded that I\ Zw\ 18 must have 
undergone an episode of star formation several hundred million years ago.

Using the same HST ultraviolet spectra, Izotov \& Thuan (1999) have 
rederived the C/O abundance ratio in both NW and SE components of I\ Zw\ 18. 
They obtained lower values of log C/O equal to --0.77 $\pm$ 0.10
for the NW component and --0.74 $\pm$ 0.09 for the SE component.
With these lower C/O ratios, I\ Zw\ 18 does not stand apart
anymore from other low-metallicity BCDs. Furthermore, the C/O ratios are in
excellent agreement with those predicted by massive star nucleosynthesis theory.
Therefore, no preceding low-mass carbon-producing stellar population
needs to be invoked, thus, supporting the original idea that I\ Zw\ 18 is
a young galaxy undergoing its first episode of star formation 
(Searle \& Sargent 1972). 
The main source of the differences ($\sim$ 0.2 dex) with the values derived by 
Garnett et al. (1997) comes from the adopted electron temperatures. 
Izotov \& Thuan (1999) use higher electron temperatures (by 1900 K and 2300 K 
respectively for the NW and SE components) as derived from recent MMT spectral 
observations in apertures which match more closely those of the HST FOS 
observations used to obtain carbon abundances. 

Thus, despite the extensive multiwavelength studies of I\ Zw\ 18, its 
evolutionary status remains controversial. Increasing evidence is accumulating, 
however, in favor of the idea that this BCD underwent its first episode of 
star formation less than 100 Myr ago. 
The first line of evidence is based on heavy element abundance ratios.
Izotov \& Thuan (1999) have studied these ratios in 
a sample of low-metallicity BCDs. They found that all galaxies with heavy 
element mass fraction $Z$ $\la$ $Z_\odot$/20, including I Zw 18, 
show constant C/O and N/O abundance ratios which can be explained by element 
production in massive stars ( $M$ $\ga$ 9 $M_\odot$ ) only. 
Intermediate-mass stars ( 3 $M_\odot$ $\la$ $M$ $<$ 9 $M_\odot$ ) in these
galaxies have not had time to die and release their C and N production.
Izotov \& Thuan (1999) put an upper limit to the age of the order of 100 Myr
for these most metal-deficient galaxies.
The chemical evidence for a young age of galaxies with $Z$ $\la$ $Z_\odot$/20
is supported by photometric and spectroscopic evidence. Thuan, Izotov \&
Lipovetsky (1997) and Papaderos et al. (1998) have argued, on the basis of
colour profiles and spectral synthesis studies, that the second most metal-deficient
galaxy known after I Zw 18, the BCD SBS 0335--052 with $Z$ $\sim$ $Z_\odot$/40,
is a young galaxy with age less than 100 Myr. Using the same techniques in
addition to colour-magnitude diagram studies, Thuan, Izotov \& Foltz (1999a)
have shown that the age of the BCD SBS 1415+437 with $Z$ $\sim$ $Z_\odot$/21
is also less than 100 Myr. 

%
%

In view of the contradictory conclusions reached by different authors based on
the same HST data set, we have decided to address anew the issue of the age
of I Zw 18 by reexamining the HST data. 
We use the archival HST WFPC2 $B$, $V$ and $R$ 
images to construct colour-magnitude diagrams for I Zw 18.
We complement the imaging data with high signal - to - noise ratio 
Multiple Mirror Telescope (MMT) spectroscopic observations. 
These data are interpreted with the help of spectral energy 
distributions (SEDs).
Spectroscopic observations are crucial for this type of analysis
as they allow to correct SEDs for contamination
by emission from ionized gas.

In Sect.\ 2 we discuss the photometric and spectroscopic data.
A determination of the 
distance to I Zw 18 and constraints on its age from $(B-V)$ vs. $V$ and 
$(V-R)$ vs. $R$ colour-magnitude diagrams are given in Sect.\ 3. 
In Sect.\ 4 age constraints inferred from the spectral energy 
distributions are discussed. Hydrodynamical age constraints from 
observations of large expanding shells
of ionized gas are considered in Sect.\ 5. We summarize our results in Sect.\ 6.

\begin{table}
\caption{Integrated photometric properties of I\ Zw\ 18}
\label{Tab1}
\begin{tabular}{lccc} \\ \hline
Parameter & Main body$^a$ & C component$^a$ & Ref. \pano\kato \\ \hline
RA(J2000)    &09$^{\rm h}$34$^{\rm m}$02.10$^{\rm s}$     &09$^{\rm h}$33$^{\rm m}$59.68$^{\rm s}$     & 1\pano \\
DEC(J2000)   &+55$^\circ$14${\rm '}$23\farcs8&+55$^\circ$14${\rm '}$42\farcs3& 1 \\
$V$ (mag)    & 16.03~\,&  19.06~\,& 1 \\
$U-B$ (mag)  &--0.96~\,& ---     & 1,2,3 \\
$B-V$ (mag)  &--0.21~\,&--0.08~\,& 1 \\
$V-R$ (mag)  &  0.24   &  0.15   & 1 \\
$V-I$ (mag)  &--0.12~\,& ---     & 1,2,3 \kato \\ \hline
\end{tabular}

\vspace*{0.5ex}
{\it References}: 1 -- Dufour et al. (1996b); 2 -- Hunter \& Thronson (1995);
3 -- this paper.

$^a$Magnitudes and colours are corrected for interstellar extinction
($A_V$ = 0.25 mag, $E(B-V)$ = 0.07 mag, $E(V-R)$ = 0.04 mag, 
Dufour et al. 1996b). $B, V, R$
magnitudes are taken from Dufour et al. (1996b), $U, I$ magnitudes are
measured in this paper within the circular aperture with radius $r$ = 
10\arcsec\ using HST observations by Hunter \& Thronson (1995).
\end{table}

\section{Observational data}

\subsection{Photometric data}

\begin{table*}
\caption{HST/WFPC2 Archival Data of I Zw 18}
\label{Tab2}
\begin{tabular}{cccccl} \\ \hline
Filter&WFPC2 quadrant&  PI  &   Epoch   & Exposure time& 
\multicolumn{1}{c}{Image root name} \\ 
      &              &      &           &     (sec)    &                 \\ 
\hline
F336W &   PC   &Hunter&29 Oct 1994&   4200      &u2cg0101t, u2cg0102t, u2cg0103t \\  
F450W &  WF3   &Dufour&03 Nov 1994&   4600      &u2f90102t, u2f90103t \\
F555W &  WF3   &Dufour&03 Nov 1994&   4600      &u2f90104t, u2f90105t \\
F702W &  WF3   &Dufour&03 Nov 1994&   5400      &u2f90101t, u2f90201t, u2f90202t \\
F814W &   PC   &Hunter&30 Oct 1994&   6600      &u2cg0301t, u2cg0302t, u2cg0303t \\ \hline
\end{tabular}


\end{table*}

   Our main goal here is to study the evolutionary status of I Zw 18,
the integrated photometric properties of which are shown in Table \ref{Tab1},
as a whole (i.e. including both the main body and the C component) by means
of colour-magnitude diagrams (CMD). To construct the CMDs,
we use archival deep $B$ (F450W), $V$ (F555W) and $R$ (F702W) HST 
observations ( PI: Dufour, GO-5434, November 1994) where both the main body
and the C component were imaged with the WFPC2 camera. Additionally,
for the measurement of the total magnitudes (Table \ref{Tab1}), 
we use $U$ (F336W) and $I$ (F814W) PC1 images of the main body (PI: Hunter, 
GO-5309, November 1994). $U$ and $I$ observations of the C component have not 
yet been done by HST.
Images reduced by the standard pipeline at the STScI were retrieved from the
HST archives. They are listed in Table \ref{Tab2}.

 The next steps of data reduction included combining all observations
with the same filter, removal of cosmic rays using the
IRAF\footnote[3]{IRAF: the Image Reduction
and Analysis Facility is distributed by the National Optical Astronomy 
Observatories, which is operated by the Association of Universities for
Research in Astronomy, In. (AURA) under cooperative agreement with the
National Science Foundation (NSF).} routine CRREJ, and correction 
for geometric distortion by producing mosaic images. The latter procedure 
introduces small photometric corrections of $\sim$ 0.02 mag for objects
near the edge of CCD chip (Holtzman et al. 1995a), like the C component,
and are even smaller for objects located far from the edge, like
the main body.

Figure\ \ref{Fig1} shows the HST WFPC2 image of I Zw 18 in the $R$ band. 
A filamentary low surface 
brightness (LSB) emission pattern extending far away from the main body of 
I\ Zw\ 18 can be seen. A number of distinct super-shells are visible within this LSB 
envelope of the main body, testimony to a large scale perturbation of 
the surrounding gaseous component and suggesting that emission by ionized gas 
contributes a significant fraction of the  integrated luminosity of the
system. This is obviously not the case for the C component
in which weak H$\alpha$ emission is barely seen and is localized in a 
compact region (Dufour et al. 1996b). 

    To construct the CMDs, we use the DAOPHOT package in IRAF. The 
point-spread function in each broad-band filter is obtained with the PSF routine 
in an interactive mode, by examining and extracting 5 -- 10 stars in 
relatively uncrowded fields. The photometry of point sources is done
with a 2-pixel radius circular aperture. Instrumental magnitudes are then
converted to magnitudes within an aperture of radius 0\farcs5, adopting 
corrections of --0.19 mag in $B$ and $V$ and --0.21 mag in $R$ (Holtzman et al.
1995a). They are finally transformed to magnitudes in the standard 
Johnson-Cousins $UBVRI$ photometric system using the prescriptions of Holtzman 
et al. (1995b).

\subsection{Spectroscopic data}

Spectroscopic observations of I\ Zw\ 18 were carried out with the MMT on 
the nights of 1997 April 29 and 30. 
Signal-to-noise ratios S/N $\ga$ 50 were reached in the continuum
of the bright central part. 
Observations were made in the blue channel of the MMT spectrograph 
using a highly optimized Loral 3072 $\times$ 1024 CCD detector. 
A 1\farcs5 $\times$ 180\arcsec\ slit was used along with a 300 groove mm$^{-1}$
grating in first order and an L-38 second-order blocking filter.
This yields a spatial resolution along the slit of 0\farcs3 pixel$^{-1}$,
a scale perpendicular to the slit of 1.9 \AA\ pixel$^{-1}$, a spectral
range 3600 -- 7500 \AA, and a spectral resolution of $\sim$ 7 \AA\ ( FWHM ).
For these observations, CCD rows were binned by a factor of 2, yielding
a final spatial sampling of 0\farcs6 pixel$^{-1}$. 
The total exposure time was 180 minutes broken up in six 
subexposures, 30 minutes each, to allow for a more effective 
cosmic-ray removal. All exposures were taken
at small airmasses ( $\la$ 1.1 -- 1.2 ), so no correction was made for 
atmospheric dispersion. The seeing during the observations was 0\farcs7 FWHM. 
The slit was oriented in the position angle P.A. = --41$^{\circ}$ to permit 
observations of the NW and SE components and the SE tip of the 
C component simultaneously. The spectrophotometric 
standard star HZ 44 was observed for flux calibration. 
Spectra of He-Ne-Ar comparison lamps were obtained before and after each
observation to provide wavelength calibration.

Data reduction of spectral observations was carried out at the NOAO
headquarters in Tucson using the IRAF software package. 
This included bias subtraction, cosmic-ray removal
and flat-field correction using exposures of a quartz incandescent
lamp. After wavelength mapping, night-sky background subtraction, 
and correcting for atmospheric extinction, each frame was
calibrated to absolute fluxes. One-dimensional spectra were
extracted by summing, without weighting, different numbers
of rows along the slit depending on the exact region of interest.
We have extracted spectra of two regions: (1) the brightest
part of the main body with size 1\farcs5 $\times$ 8\farcs5 centered on the NW
component and (2) the southeastern tip of the C component
within an aperture 1\farcs5 $\times$ 3\arcsec. The observed and 
extinction-corrected emission-line intensities in the main body 
of I\ Zw\ 18 are listed in Table \ref{Tab3}.

The ionic and elemental abundances have been derived following the procedure by 
Izotov et al. (1994, 1997c). The extinction coefficient $C$(H$\beta$) and the
absorption equivalent width $EW$(abs) for the hydrogen lines obtained
by an iterative procedure are included in Table \ref{Tab3} together with the 
observed flux $F({\rm H}\beta)$
and the equivalent width $EW({\rm H}\beta)$ of the H$\beta$ emission line.
The electron temperature $T_e$(O III) was determined
from the [O\,III] $\lambda$4363/ ($\lambda$4959 + $\lambda$5007) flux
ratio and the electron number density $N_e$(S II) from the 
[S\,II] $\lambda$6717/$\lambda$6731 flux ratio.  The ionic and elemental
abundances are shown in Table \ref{Tab4} together with ionization correction 
factors (ICFs). 
They are in good agreement with the abundances derived by Skillman \&
Kennicutt (1993), Izotov \& Thuan (1998), V\'ilchez \& Iglesias-P\'aramo 
(1998) and Izotov et al. (1999).

%
\begin{table}
\caption{Emission line intensities in the main body of I Zw 18}
\label{Tab3}
\begin{tabular}{lcc} \hline
Ion   &$F$($\lambda$)/$F$(H$\beta$)
&$I$($\lambda$)/$I$(H$\beta$)
 \\ \hline
 3727\ [O II]        &0.3473$\pm$0.0035& 0.3547$\pm$0.0038 \\
 3835\ H9            &0.0195$\pm$0.0021& 0.0971$\pm$0.0141 \\
 3868\ [Ne III]      &0.1519$\pm$0.0028& 0.1537$\pm$0.0029 \\
 3889\ He I + H8     &0.1191$\pm$0.0026& 0.1956$\pm$0.0052 \\
 3968\ [Ne III] + H7 &0.1446$\pm$0.0026& 0.2170$\pm$0.0048 \\
 4101\ H$\delta$     &0.2072$\pm$0.0027& 0.2720$\pm$0.0042 \\
 4340\ H$\gamma$     &0.4204$\pm$0.0034& 0.4701$\pm$0.0043 \\
 4363\ [O III]       &0.0639$\pm$0.0022& 0.0628$\pm$0.0022 \\
 4471\ He I          &0.0263$\pm$0.0019& 0.0257$\pm$0.0019 \\
 4686\ He II         &0.0249$\pm$0.0019& 0.0240$\pm$0.0019 \\
 4861\ H$\beta$      &1.0000$\pm$0.0053& 1.0000$\pm$0.0056 \\
 4959\ [O III]       &0.6927$\pm$0.0042& 0.6605$\pm$0.0042 \\
 5007\ [O III]       &2.0835$\pm$0.0092& 1.9822$\pm$0.0092 \\
 5876\ He I          &0.0838$\pm$0.0017& 0.0768$\pm$0.0017 \\
 6300\ [O I]         &0.0090$\pm$0.0014& 0.0081$\pm$0.0013 \\
 6312\ [S III]       &0.0091$\pm$0.0012& 0.0082$\pm$0.0011 \\
 6563\ H$\alpha$     &3.0488$\pm$0.0126& 2.7450$\pm$0.0129 \\
 6678\ He I          &0.0291$\pm$0.0012& 0.0259$\pm$0.0012 \\
 6717\ [S II]        &0.0334$\pm$0.0014& 0.0297$\pm$0.0013 \\
 6731\ [S II]        &0.0216$\pm$0.0014& 0.0192$\pm$0.0013 \\
 7065\ He I          &0.0251$\pm$0.0011& 0.0221$\pm$0.0011 \\
 7135\ [Ar III]      &0.0195$\pm$0.0011& 0.0172$\pm$0.0011 \\ \\
 $C$(H$\beta$) dex    &\multicolumn {2}{c}{0.090$\pm$0.005} \\
 $F$(H$\beta$)$^a$ &\multicolumn {2}{c}{ 3.50$\pm$0.02} \\
 $EW$(H$\beta$)\ \AA &\multicolumn {2}{c}{68$\pm$5} \\
 $EW$(abs)\ \AA      &\multicolumn {2}{c}{3.0$\pm$0.1} \\ \hline 
\end{tabular}

$^a$in units of 10$^{-14}$ ergs\ s$^{-1}$cm$^{-2}$.
\end{table}

%
\begin{table}
\caption{Heavy Element Abundances in the main body of I Zw 18}
\label{Tab4}
\begin{center}
\begin{tabular}{lc} \hline
Parameter&Value \\ \hline
$T_e$(O III)(K)                     &19300$\pm$400      \\
$T_e$(O II)(K)                      &15500$\pm$300      \\
$T_e$(S III)(K)                     &17700$\pm$300      \\
$N_e$(S II)(cm$^{-3}$)              &    10$\pm$10      \\ \\
O$^+$/H$^+$($\times$10$^5$)         &0.280$\pm$0.014    \\
O$^{++}$/H$^+$($\times$10$^5$)      &1.221$\pm$0.058    \\
O$^{+3}$/H$^+$($\times$10$^5$)      &0.042$\pm$0.004    \\
O/H($\times$10$^5$)                 &1.544$\pm$0.060    \\
12 + log(O/H)                       &7.189$\pm$0.017    \\ \\
Ne$^{++}$/H$^+$($\times$10$^5$)     &0.194$\pm$0.010    \\
ICF(Ne)                             &1.26\,~~~~~~~~~~   \\ 
log(Ne/O)                           &--0.799$\pm$0.033~~\\ \\
S$^+$/H$^+$($\times$10$^7$)         &0.457$\pm$0.022    \\
S$^{++}$/H$^+$($\times$10$^7$)      &2.619$\pm$0.377    \\
ICF(S)                              &1.67\,~~~~~~~~~~   \\ 
log(S/O)                            &--1.478$\pm$0.045~~\\ \\
Ar$^{++}$/H$^+$($\times$10$^7$)     &0.473$\pm$0.030    \\
ICF(Ar)                             &2.00\,~~~~~~~~~~   \\
log(Ar/O)                           &--2.213$\pm$0.026~~\\ \hline
\end{tabular}
\end{center}
\end{table}

\section{Stellar population ages from colour-magnitude diagrams}
%

The superior spatial resolution of HST WFPC2 images combined with the
proximity of I\ Zw\ 18 
permits to resolve individual bright stars and study stellar populations in 
this galaxy by means of colour - magnitude diagrams. 
Such a study has been done by Hunter \& Thronson (1995) for the main body
in the $U$, $V$ and $I$ bands. Their photometry shows a broad main sequence
of massive stars and blue and red supergiants. The NW component contains the 
brightest and reddest, presumably most evolved stars, spanning a range 2 -- 
5 Myr in age. The stars in the SE component are likely to be even younger.

Dufour et al. (1996b) were able to resolve the C component into stars and found
it to be older than the NW and SE components. On the basis of 
$(B-V)$ vs. $V$ colour-magnitude diagrams, they 
concluded that star formation started in the main body several tens
of Myr ago and in the C component several hundred of Myr ago. 
In this section we re-analyze the observations by Dufour et al. (1996b) and
re-examine their conclusions concerning the age of the stellar content of 
I\ Zw\ 18. In order to compare observed to theoretical CMDs and derive ages, a 
precise distance to I Zw 18 is needed, which we discuss next.

\subsection{The distance to I Zw 18}

For the nearest galaxies, distances can be derived from colour-magnitude 
diagrams themselves by measuring the apparent magnitude of the tip
of the red giant branch clump (e.g. Schulte-Ladbeck, Crone \& Hopp 1998; Lynds 
et al. 1998). However, red giants are too faint to be seen in more distant 
galaxies (such as I Zw 18), and other methods must be used. 
Any such distance determination should always be checked for consistency,
i.e. galaxy properties derived from CMD analysis such as the age of the
stellar populations or the luminosities of the brightest stars should be
compatible with other known observed characteristics of the galaxy.

 A distance of 10 Mpc to I Zw 18 has generally been adopted by previous 
authors (Hunter \& Thronson 1995, Dufour et al. 1996b and Aloisi et al. 1999).
This assumes that the observed heliocentric radial velocity of the galaxy 
$\sim$ 740 km s$^{-1}$ is pure Hubble flow velocity and a Hubble constant
$H_0$ = 75 km s$^{-1}$ Mpc$^{-1}$. Adopting this distance would lead to a
conflict with the well-observed ionization state of I Zw 18.
At 10 Mpc the brightest stars observed in the main body and in the C component 
have absolute $V$ magnitudes fainter than --8 and --6 mag respectively (Hunter 
\& Thronson 1995; Dufour et al. 1996b; Aloisi et al. 1999). 
If that is the case, comparison with evolutionary tracks implies that the most 
massive stars 
in the main body and C component would have masses less than 15 $M_\odot$ and 
9 $M_\odot$ respectively (Dufour et al. 1996b; Aloisi et al. 1999). 
 According to Vacca (1994), O stars have masses exceeding 13 $M_\odot$.
Thus, only very late O stars would be present in the main body. This conclusion
is in severe contradiction with the observed ionization state of I Zw 18.
Indeed, the equivalent width of the H$\beta$ emission line expected from a
stellar population having an upper mass limit of 15 $M_\odot$ is $<$ 10 \AA, 
while the observed H$\beta$ equivalent widths in the NW and SE components lie in
the range 60 -- 130 \AA, implying the presence of stars with masses
$\ga$ 40 -- 50 $M_\odot$. If the upper stellar mass limit of 9 $M_\odot$ 
derived by Dufour et al. (1996b) and Aloisi et al. (1999) for the C component is
correct, then ionized gas should not be present in this component because of
the lack of O and early B stars. 
But H$\alpha$ and H$\beta$ are clearly observed (Dufour et al. 1996ab; 
Izotov \& Thuan 1998; van Zee et al. 1998; this paper) in the C component. 
Izotov \& Thuan (1998) derived $EW$(H$\beta$) = 6 \AA, which, after correction
for underlying stellar absorption, would result in a
value as high as 10 \AA. Thus, late O and early B stars with masses as high
as 15 $M_\odot$ must be postulated in the C component. 
We argue therefore that
the stellar absolute magnitudes derived by Dufour et al. (1996b) and Aloisi et 
al. (1999) from their CMDs are too faint because they
are based on too small an adopted distance. 

%
\begin{figure}
\vspace{13.0cm}
\includegraphics{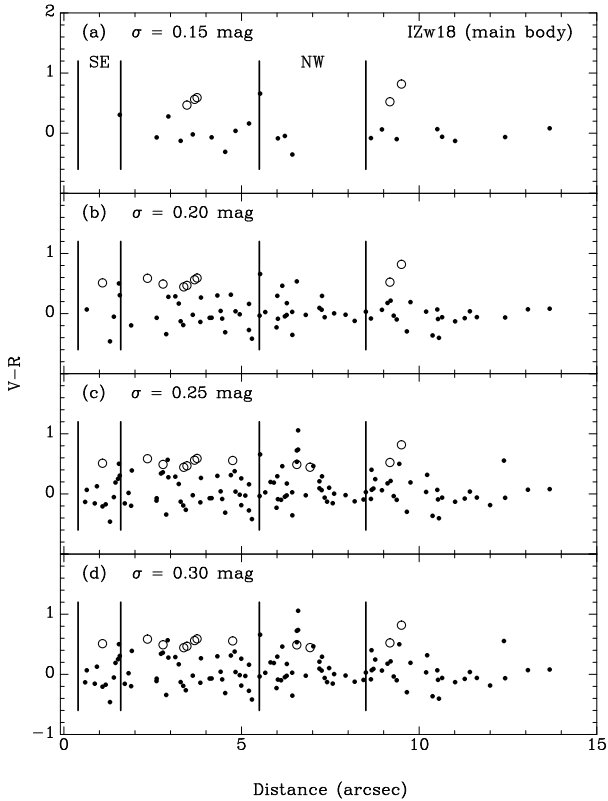}
\vspace{-0.7cm}
\caption[]{\label{Fig3}
Colours of point sources in the main body as a function of angular 
distance from the southeastern tip of the SE component, for different
upper limits of the photometric uncertainty $\sigma$. The boundaries of the 
SE and NW components are shown by vertical lines. Faint red stars 
with $R$ $\ga$ 25 mag and ($V-R$) $\ga$ 0.4 mag are shown by open circles. }
\end{figure}

The real distance to I Zw 18 appears to be considerably larger. Correction 
of the I Zw 18 heliocentric radial velocity to the centroid of the Local 
Group and for a Virgocentric infall motion of 300 km s$^{-1}$ 
( Kraan-Korteweg 1986 ) gives a velocity of 1114 km s$^{-1}$.
If the Hubble constant is in the currently accepted range of 55 -- 70 
km s$^{-1}$ Mpc$^{-1}$ this would give a distance between $\sim$ 16 and 
$\sim$ 20 Mpc for I Zw 18.

Because of uncertainties introduced by the statistical nature of the
Virgocentric infall correction and those in the value of the Hubble constant,
we prefer to base our estimate of the distance to I Zw 18 on two firm 
observational results: 1) the presence of ionized gas in the C component 
and 2) the presence of WR stars in the NW component. Concerning the
ionized gas in the C component, the distance should be increased to a value
of $\sim$ 20 Mpc. Increasing the distance by a factor of 2 would make the most
massive stars more luminous by a factor of 4 and push the mass upper limit to
$\sim$ 15 $M_\odot$. These more massive stars would then provide enough
ionizing photons to account for the observed emission lines in the C component.
Concerning the WR stars, they have been seen both spectroscopically (Izotov et 
al. 1997a; Legrand et al. 1997) and photometrically 
(Hunter \& Thronson 1995; de Mello et al. 1998) in the NW component of I Zw 18, 
in the region where the brightest post-main-sequence stars are located. 
The existence of WR stars implies 
the presence of very massive stars in the NW component. 
De Mello et al. (1998) using Geneva stellar evolutionary tracks for
massive stars with enhanced stellar wind and with heavy element mass fraction 
$Z$ = 0.0004 found the minimum initial mass $M_{min}$ for stars evolving 
to WR stars to be $\sim$ 90 $M_\odot$ and that the WR stage is very short-lived, being only $\sim$ 0.8 Myr, if the instantaneous burst model is adopted. 
The models used by de Mello et al. (1998) do not take rotation into account,
which may decrease $M_{min}$, but probably by not more
than a factor of 1.5 (Langer \& Heger 1998; Meynet 1998). 
Thus the observation of WR stars in the NW component implies that 
post-main-sequence massive stars with $M$ $\ga$ 40 -- 60 
$M_\odot$ and with lifetimes $\la$ 3 -- 5 Myr (Fagotto et al. 1994;
Meynet et al. 1994) must be present in I Zw 18.
This short time scale is in excellent agreement with the age of 5 Myr
derived from the equivalent width of H$\beta$ in the main body 
(Table \ref{Tab3}), using the calibration by Schaerer \& Vacca (1998). 
To accomodate this small time scale, we are forced again to increase the
distance of I Zw 18 to $\sim$ 20 Mpc. Increasing the distance would increase
the luminosity of all stars and the location of the brightest stars in 
the CMD of the main body can be accounted for by isochrones 
with age as short as $\sim$ 5 Myr.

    In summary, three different lines of argument --- a non Hubble flow
velocity component, the ionization state of I Zw 18 and the presence of WR
stars in its NW component --- have led us to conclude
that I Zw 18 is twice as distant as thought before. We shall thus adopt 
a distance of 20 Mpc to I Zw 18. At this distance, 1\arcsec\ = 97 pc.

%
%
\begin{figure*}
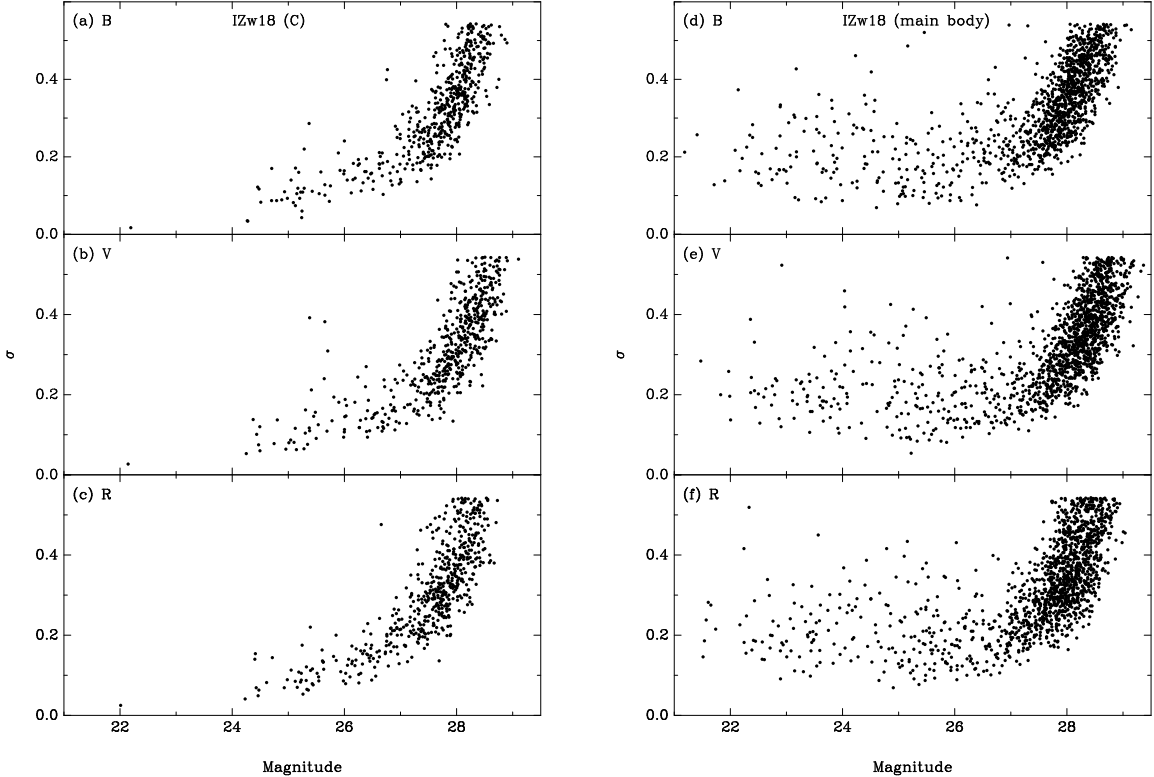

\vspace{13.0cm}
\includegraphics{fig3a.ps}
\includegraphics{fig3b.ps}
\vspace{-0.7cm}
\caption[]{\label{Fig2}
Photometric uncertainties $\sigma$ in the determination of apparent magnitudes of 
point sources in the C component (left) and in the main body (right)
of I\ Zw\ 18. Note the much larger errors at a fixed magnitude in the main body 
as compared to the C component. The larger errors are due to the contamination 
of ionized gas and severe stellar crowding.}
\end{figure*}

\subsection{The resolved component and colour-magnitude diagrams}

In Fig. \ref{Fig3} we show the spatial distribution of the ($V-R$) colours of point sources 
along the main body for increasing photometric uncertainties $\sigma$. 
Panels (a) to (d) show CMD data points
with successively larger photometric errors, but smaller than the $\sigma$ value
given in each panel. The origin is taken to be at the south-eastern tip of the
main body. The boundaries of the SE and NW components are shown by vertical
lines. In the most crowded fields of the SE and NW starburst components, 
$V$ and $R$
magnitudes are measured with a precision better than 0.15 mag only for very few
stars (Fig. \ref{Fig3}a). However, if the upper limits are greater than
$\ge$ 0.25 mag, then almost all stars in these two components are recovered
as suggested by the similarity of the distributions of stars in panels (c) and 
(d).
Figure \ref{Fig2} shows the distribution of photometric errors as a function of 
apparent $B$, $V$ and $R$ stellar magnitudes for the C component 
(Figs. \ref{Fig2}a -- \ref{Fig2}c) 
and the main body (Figs. \ref{Fig2}d -- \ref{Fig2}f).
The photometric errors in the $V$ band at a fixed magnitude are 
similar to those derived by Hunter \& Thronson (1995). In the main body, 
the errors remain rather large even for the brightest stars 
( $\sigma$ $\sim$ 0.2 mag at $V$ $\sim$ 22 mag ). 
This is due to the spatially varying background in the highly crowded 
region of the main body. These photometric errors are especially large
at the faint magnitudes of red 
evolved stars ( $\sigma$ $\sim$ 0.4 mag at $V$ $\sim$ 28 mag ). By contrast, 
the photometric errors in all filters are lower for 
the sources in the C component, being $\sim$ 0.1 mag at $V$ $\sim$ 25 mag and
increasing to $\sim$ 0.3 mag at $V$ $\sim$ 28 mag, because stellar crowding and 
contamination of the stellar background by the ionized gas emission are smaller. 

%
%
%
\begin{figure}[t]
\vspace{13.0cm}
\includegraphics{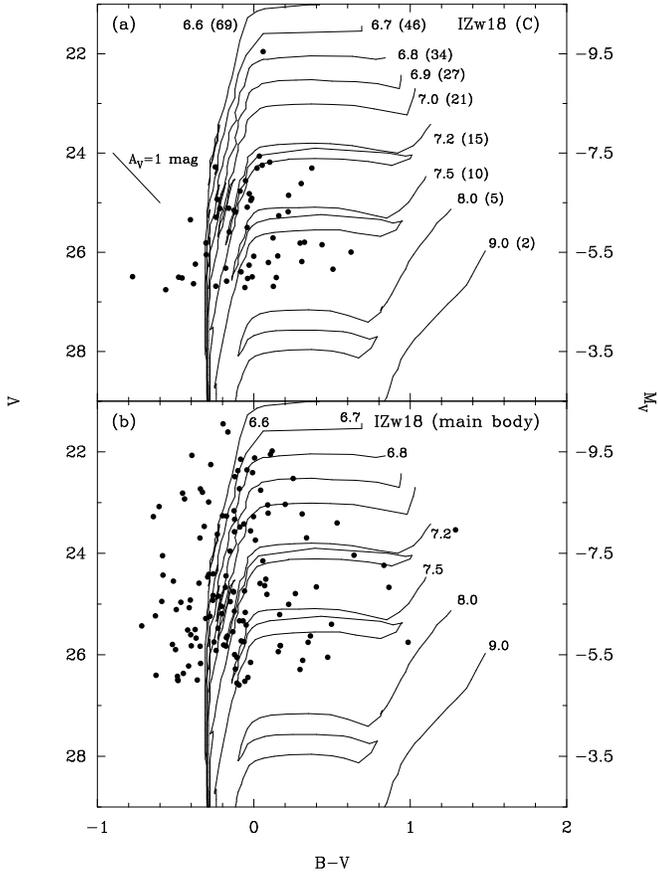}
\vspace{-0.7cm}
\caption[]{\label{Fig4}
$(B-V)$ vs. $V$ colour-magnitude diagram for point sources in I Zw 18. 
Theoretical isochrones from Bertelli et al. (1994) for a stellar population with 
a heavy element mass fraction $Z$ = 0.0004 are shown by solid lines 
labeled by the logarithm of age in yr and, in parentheses by the maximum 
stellar mass corresponding to each isochrone. A distance of 20 Mpc is adopted.}
\end{figure}
%

%
%
\begin{figure}[t]
\vspace{13.0cm}
\includegraphics{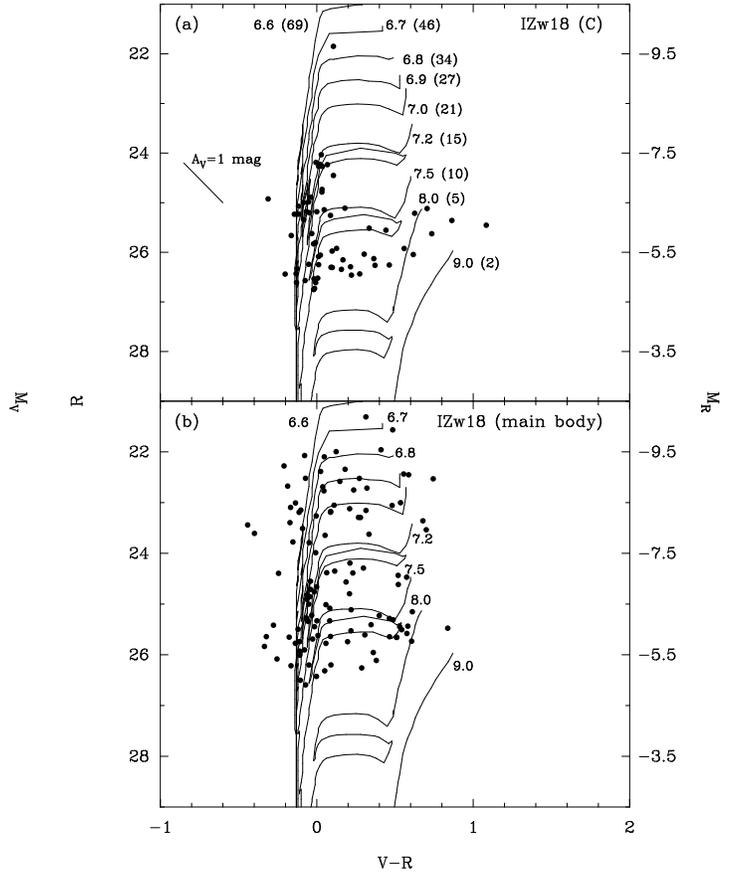}
\vspace{-0.7cm}
\caption[]{\label{Fig5}
$(V-R)$ vs. $R$ colour-magnitude diagram for point sources in I Zw 18. 
Theoretical isochrones from Bertelli et al. (1994) for a stellar population 
with a heavy element mass fraction of $Z$ = 0.0004 are shown by solid curves
and they labeled as in Fig. \ref{Fig4}. A distance of 20 Mpc is adopted.}
\end{figure}

    In the following CMD analysis, we shall consider only point sources which 
are brighter than 27 mag in each band and those for which the photometric errors 
do not exceed 0.25 mag. For the C component, each of these two conditions 
selects out nearly the same point sources, as 27 mag sources have a mean error
of $\sim$ 0.25 mag. However, in the main body,
some bright sources are rejected because they have photometric
errors larger than 0.25 mag (Fig. \ref{Fig2}).

%

The correction for internal extinction poses a problem. The extinction
for the main body can be estimated from the optical emission-line spectrum. 
However, different measurements give somewhat different values for the 
extinction. Dufour et al. (1996b) have adopted a value $A_V$ = 0.25 mag based
on the spectroscopic observations of Skillman \& Kennicutt (1993).
The value of $C$(H$\beta$) = 0.09 derived in this paper (Table \ref{Tab3}) 
for the main body translates into a mean $A_V$ = 0.19 mag. 
The extinction may differ in the NW and SE components and it is not possible 
to estimate it in the C component because Balmer hydrogen emission lines in its
spectrum are not detected with high enough signal-to-noise ratio.

It is worth noting that recent studies reveal that dust formation may 
proceed very efficiently in a low-metallicity starburst environment and 
cause a highly inhomogeneous extinction pattern.
Thuan, Sauvage \& Madden (1999b) have analyzed ISO
observations of the second most metal - deficient galaxy known, SBS 0335--052, 
and found that an intrinsic extinction of up to $A_V$ = 19 -- 21 mag is 
required around some stellar complexes to account for the galaxy's spectral 
energy distribution in the mid-infrared. The mean extinction in SBS 
0335--052 derived from the optical spectroscopic observations is much lower 
( $A_V$ $\sim$ 0.5 mag, Izotov et al. 1997b), although
it varies significantly along the slit. Thuan et al. (1999b) found that as much
as 3/4 of the massive stars in SBS 0335--052 may be hidden by dust.

Keeping in mind these findings, we adopt for simplicity and for lack of more 
information a spatially constant extinction of 
$A_V$ = 0.19 mag, $E(B-V)$ = 0.06 mag and $E(V-R)$ = 0.04 mag 
for the main body of I\ Zw\ 18 (Table \ref{Tab3}), while 
the extinction for the C component is taken to be zero. 
An inhomogeneous and locally much higher
extinction than the value adopted above will evidently 
result in a larger dispersion in the CMDs and in increased colour
indices, i.e. in an overestimate of the age for I\ Zw\ 18.

We show the $(B-V)$ vs. $V$ diagram for the
C component in Fig. \ref{Fig4}a and that for the main body in Fig. \ref{Fig4}b. 
Solid lines represent theoretical isochrones (Bertelli et al. 1994), for a 
heavy element mass fraction $Z$ = 0.0004, labeled according to 
the logarithm of age. The largest stellar mass 
associated with each isochrone is also given in parentheses.
It may be seen that there exists a well-defined main sequence in the 
C component with a turn-off indicative of an age of $\sim$ 15 Myr. 
There is a very bright point source with $V$ $\sim$ 22 mag, which may be
in fact a compact star cluster as suggested by Dufour et al. (1996b). Although
no red sources with ($B-V$) $\ga$ 1 are seen, there is a group of relatively
redder points with 0.2 $\la$ ($B-V$) $\la$ 0.6 in Fig. \ref{Fig4}a.
Because of their faint magnitudes ( $V$ $\sim$ 26 ), some of the
red sources may be attributed to photometric uncertainties ( $\sigma$ $\sim$
0.2 mag at $B$ and $V$ $\sim$ 26, resulting in
$\sigma$($B-V$) $\sim$ 0.3 mag, Fig. \ref{Fig2} ). The fact that there are also
points that scatter to the blue at that faint magnitude would support that
hypothesis. However, even if we do accept that all these faint red stars are
real, comparison with theoretical isochrones says that their age cannot exceed
100 Myr.

   As for the main body, the main sequence turn-off at $V$ $\sim$ 22 mag
implies an age of $\sim$ 5 Myr ( Fig. \ref{Fig4}b ). 
Red stars ( ($B-V$) $\ga$ 0.2 ) are seen to possess a large range of absolute 
magnitudes ( $\sim$ 3 -- 4 mag ) implying that star formation 
has been undergoing in the main body in the last $\sim$ 15 -- 30 Myr.
This conclusion is essentially the same as that by Dufour et al. (1996b)
except for differences introduced by the larger distance to I Zw 18 which 
results in smaller ages. We note that the spread of the points
in the $(B-V)$ vs. $V$ diagram of the main body (Fig. \ref{Fig4}b) is larger 
than the one in the CMD of the C component. It is probably not only due to 
evolutionary effects, but also to larger photometric uncertainties 
in the main body because of more crowding and contamination from gaseous emission 
as discussed above (cf. Fig. \ref{Fig2}).

$(V-R)$ vs. $R$ CMD diagrams for the C component and the main body
along with theoretical isochrones from Bertelli et al. (1994) for 
$Z$ = 0.0004 (solid lines) are shown in Fig. \ref{Fig5}. 
As in Fig. \ref{Fig4}, the logarithm of age and the maximum stellar mass 
are given for each isochrone.

Figure \ref{Fig5}a shows a well-defined main-sequence for component C
corresponding to an age of $\sim$ 15 Myr. Several red and faint 
($V$ $\ga$ 25 mag) stars with 
$(V-R)$ $\la$ 1.1 mag are seen. They probably are red supergiants and/or 
massive young luminous asymptotic giant branch (AGB) stars, $\sim$ 2 mag 
brighter than the older AGB stars observed in the BCD VII Zw 403 (Lynds et al. 
1998; Schulte-Ladbeck et al. 1998) and Local Group galaxies ( e.g. Gallart, 
Aparicio \& V\'ilchez 1996 ). The location of these 
red stars in the CMD suggests an age of  $\la$ 100 Myr. We conclude that star 
formation in the C component probably began $\sim$ 100 Myr ago 
and finished $\sim$ 15 Myr ago.

The spread of points in the main body ( Fig. \ref{Fig5}b ) is similar to that 
in the $(B-V)$ vs. $V$ CMD (Fig. \ref{Fig4}). Again, red stars in the main body 
span a range by 3 -- 4 magnitudes in brightness suggesting that star formation 
has occurred during the last $\sim$ 15 -- 30 Myr.
The age of $\sim$ 5 Myr derived for the brightest stars in the main body is in
agreement with that inferred from spectral observations ( see Sect. 4 ). 

   Our derived upper age limit of $\sim$ 100 Myr for I Zw 18 differs by a whole
order of magnitude from the age of up to 1 Gyr derived by Aloisi et al. (1999) 
in their CMD analysis of the archival $B$, $V$, $I$ HST data. This large age 
difference comes mostly from the doubling of the distance to I Zw 18.
Aloisi et al. (1999) found an appreciable number of faint red sources with 
$V$ = 26 -- 27 mag and ($V-I$) = 1 -- 2 mag. With our new distance
of 20 Mpc, these sources have absolute $I$ magnitude of --6.5 mag, 
or 2 mag brighter than the absolute $I$ magnitude of older and fainter AGB stars
detected in the BCD VII Zw 403 (Lynds et al. 1998; Schulte-Ladbeck et al.
1998). Because of their high luminosities, these red stars cannot be
interpreted as old AGB stars as proposed by Aloisi et al. (1999). Rather, it is 
more likely that they are young stars. At the new distance, the location in the 
($V-I$) vs. $V$ diagram of the faint red stars seen by Aloisi et al. (1999) is 
consistent with an age of $\sim$ 50 Myr if the Geneva tracks are used, and of 
$\la$ 100 Myr if the Padova tracks are used. These red stars with $R$ = 25 -- 26 mag and 
($V-R$) $\sim$ 0.6 are also seen by us in the ($V-R$) vs $R$
CMD (Fig. \ref{Fig5}b) and their age is again consistent
with $\la$ 50 Myr as inferred from the Padova isochrones. 
Aloisi et al. (1999) restricted their CMDs to points with photometric errors smaller than 
$\sigma$ = 0.2 mag. With such a cut, those authors found
that the faint red stars are located mainly in the southeastern part of
the main body. Adopting the same cut, we come to the same conclusion as can
be seen in Figure \ref{Fig3}b, where
these stars are shown by open circles. They are seen to be located 
mainly in the relatively uncrowded
region between the SE and NW components. However, if a slightly 
larger cut of $\sigma$ = 0.25
mag is used instead, the distribution of faint red stars in the main body 
becomes more uniform (Figure \ref{Fig3}c). The ages of the faint red stars 
of $\sim$ 100 Myr derived here, should be considered as upper limits. If these 
sources are subject to local high extinction, their ages will be decreased.

%
%
\begin{figure}[t]
\vspace{14.0cm}
\includegraphics{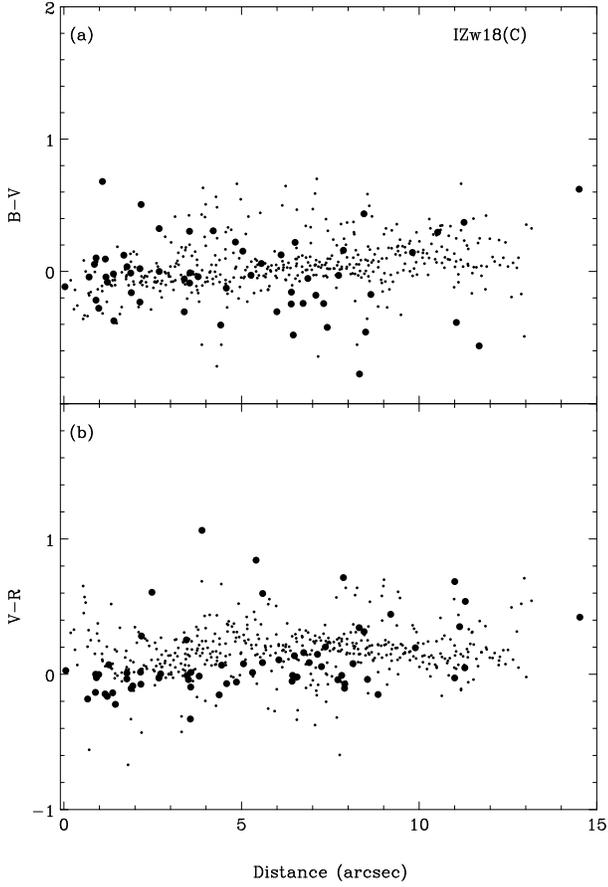}
\vspace{-0.7cm}
\caption[]{\label{Fig6}
Colours of point sources (filled circles) and of the diffuse unresolved
low-surface-brightness (LSB) emission (dots) in the C component as function of 
angular distance from the southeastern tip of the C component.
A slight $(V-R)$ colour gradient for resolved sources is present
which suggests that bright stars in the northwestern part of the 
C component are slightly redder compared to stars in the southeastern part.}
\end{figure}

    In summary, the upper ages derived for point sources 
from colour-magnitude diagrams in both the main body and C component 
of I Zw 18 do not exceed 100 Myr.

Figure \ref{Fig6} shows the radial colour distribution of point sources 
along the body of the C component with the origin 
taken to be at the southeastern tip. 
While the ($B-V$) colour does not show a gradient (Fig. \ref{Fig6}a), 
there is a weak trend for the $(V-R)$ colour of stars to become redder
away from the southeastern tip towards the northwestern tip of the C component.
This trend was also discussed by Dufour et al. (1996b) and may 
reflect propagating star formation in the C component from the NW to the SE. 



The dots in Fig.\ \ref{Fig6} show the colours of the unresolved stellar 
continuum of the C component down to a surface brightness limit 
of 25 $R$ \sbu, averaged within regions 
of size 0\farcs 4 $\times$ 0\farcs 4. 
The mean colours of the diffuse component were determined to be
$<\!B-V\!>$ = +0.05 mag and $<\!V-R\!>$ = +0.16 mag with a standard deviation 
of 0.01 mag around the mean. 

A colour gradient 
in the diffuse stellar continuum is more evident in ($B-V$) than in ($V-R$), amounting to (0.15$\pm$0.03) mag kpc$^{-1}$ and (0.05$\pm$0.025) 
mag kpc$^{-1}$, respectively.
The diffuse stellar continuum distribution follows very closely that 
of the resolved point sources, although a small difference in the
$(V-R)$ colour of the resolved and unresolved components exists in the 
southeastern part. These results imply that, contrary to the majority of BCDs
(e.g. Papaderos et al. 1996a), in which star-forming regions are immersed in
an extended and much older stellar LSB envelope, the blue unresolved stellar 
continuum in the C component is of comparable age or even formed coevally with 
the population of resolved sources discussed in this section.
The integrated photometric properties of the unresolved components 
in both the main body and C component are listed in Table \ref{Tab5}.
They are derived from the total luminosity of each component after the 
summed up emission of point-like sources has been subtracted. 
Quantities listed in Table \ref{Tab5} referring to the main body of
I\ Zw\ 18 are corrected for intrinsic extinction.

Note that while the emission of the resolved component in the
main body of I\ Zw\ 18 is contributed by stellar sources, 
the unresolved emission includes both stellar and gaseous contributions.
It follows from Table \ref{Tab5} that most of the light comes from 
the unresolved stellar component.
The contribution of resolved sources
to the total light of both the main body and C component is $\la 25$\%.
%
\begin{table}
\caption{Photometric properties of resolved and diffuse emission in I\ Zw\ 18}
\label{Tab5}
\begin{tabular}{lcc} \\ \hline
Parameter    &  Main body$^a$& C component$^b$ \\ \hline
\multicolumn{3}{c}{a) Total} \\ 
$V$ (mag)    &  16.05~\,& 19.20~\,  \\ 
$B-V$ (mag)  &--0.18~\,&  0.00  \\
$V-R$ (mag)  &  0.20   &  0.14  \\ 
\multicolumn{3}{c}{b) Resolved population} \\ 
$V$ (mag)    &  18.35~\,& 20.69~\,  \\ 
$B-V$ (mag)  &--0.41~\,&--0.03~\,  \\
$V-R$ (mag)  &  0.13   &  0.11  \\ 
\multicolumn{3}{c}{c) Unresolved population} \\ 
$V$ (mag)    &  16.16~\,& 19.46~\,  \\ 
$B-V$ (mag)  &--0.16~\,&  0.01  \\
$V-R$ (mag)  &  0.21   &  0.16  \\ \hline
\end{tabular}

\vspace*{0.4ex}
$^a$ Corrected for extinction as derived from spectroscopic
observations ($A_V$ = 0.19 mag, $E(B-V)$ = 0.06 mag, $E(V-R)$ = 0.04 mag). 
The measurements have been done within a circular aperture of 
10\arcsec\ in radius. \\[0.5ex]
$^b$ Not corrected for extinction.
\end{table}

%
\section{Synthetic spectral energy distribution}
%

As discussed before, the stellar emission in the main body of 
I\ Zw\ 18 is strongly contaminated by emission of ionized gas from supergiant
\h2\ regions. Therefore, to derive the BCD's stellar populations age, 
synthetic spectral energy distributions
which include both stellar and ionized gaseous emission need to
be constructed. By contrast, the contamination of the light of the C component 
by gaseous emission is small, and photometric and spectral data 
give us direct information on its stellar populations. 

To analyze the stellar populations in the young ionizing clusters
of the main body, we use SEDs calculated by 
Schaerer (1998, private communication) for a heavy element mass fraction of 
$Z$ = $Z_\odot$/20 (SEDs with lower metallicity are not available) 
and ages in the range of $t$ = 2 -- 10 Myr. As for the C
component, the absence of strong ionized gas emission implies that its stellar 
population is older than 10 Myr.
We have therefore calculated a grid of SEDs for stellar populations with ages
between 4 Myr and 20 Gyr and heavy element abundances $Z/Z_\odot$ = 10$^{-2}$, 
using the stellar isochrones of Bertelli et al. (1994) and the compilation of
stellar atmosphere models of Lejeune, Cuisinier \& Buser (1998). An initial 
mass function (IMF) with a Salpeter slope equal to --2.35, an upper mass limit
of 100 $M_\odot$ and 
a lower mass limit of 0.6 $M_\odot$ are adopted. The {\it observed} gaseous 
spectral energy distribution is then added to the calculated stellar spectral 
energy distribution, its contribution being determined by
the ratio of the observed equivalent width of the H$\beta$ emission line
to the one expected for pure gaseous emission. To calculate the gaseous 
continuum spectral energy distribution, the observed H$\beta$ flux and the 
electron temperature have been derived from the spectrum at 
each point along the slit. The contribution of bound-free, free-free and 
two-photon continuum emission has been taken into account
for the spectral range from 0 to 5 $\mu$m (Aller 1984; Ferland 1980).
Emission lines are superposed on top of the gaseous continuum SED with 
intensities derived from spectra in the spectral range 
$\lambda$3700 -- 7500 \AA. Outside this range, the intensities
of emission lines (mainly hydrogen lines) have been calculated
from the extinction-corrected intensity of H$\beta$.

\subsection{The main body}

\begin{figure}[t]
\vspace{9.0cm}
\includegraphics{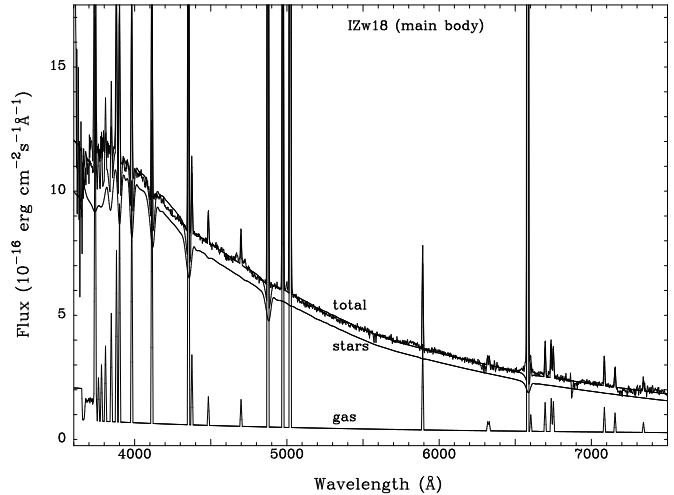}
\vspace{-0.7cm}
\caption[]{\label{Fig12}
MMT spectrum of the main body of I\ Zw\ 18 on which is superposed a synthetic 
continuum including both gaseous and stellar emission for a composite
stellar population composed of a young population with age 2 Myr 
(40 \% of total mass) and an older one with age 5 Myr (60 \%
of total mass) (thick line). The observed spectral energy
distribution is corrected for an extinction $A_V$ = 0.19 mag, derived 
from the Balmer hydrogen emission lines decrement. }
\end{figure}

Fig.\ \ref{Fig12} shows the observed spectrum of the main body ( NW + SE components ) 
along with the synthesized gaseous and stellar 
continuum for a composite stellar population of ages 2 and 5 Myr 
contributing respectively 40 \% and 60 \% of the total mass, in the
approximation of an instantaneous burst of star formation.
The line intensities, electron temperature, electron number density and 
heavy element abundances are taken from Tables \ref{Tab3} and \ref{Tab4}.
Comparison with synthesized SEDs of different ages shows that there is good
agreement with the observed SED only for an age of 5 Myr. However, the
synthesized SED is systematically below the observed SED in the blue part 
of the spectrum with $\lambda$ $\la$ 4000 \AA. Hence, an additional younger 
stellar population with age $\sim$ 2 Myr is required. We conclude that the 
emission of the main body is dominated by a very young 
stellar population with age of $\la$ 5 Myr. This age is 
considerably shorter than the one
of 15 -- 20 Myr derived by Aloisi et al. (1999) for their postulated second
burst of star formation, and much shorter still than the 
30 Myr -- 1 Gyr time scale for the first episode of star formation.

The contribution of ionized gas to the SED of star-forming regions in 
the main body is not as large in I Zw 18 as in SBS 0335--052 
(Papaderos et al. 1998), as can be seen by the equivalent width of H$\beta$ 
and the observed intensities of other emission lines relative to H$\beta$.
In the brightest part of the main body, the equivalent width of the
H$\beta$ emission line is only 68 \AA\ (Table \ref{Tab4}) as compared to 
$\ga$ 200 \AA\ in SBS 0335--052.

%
%

\subsection{The C component}

%
%
\begin{figure}[t]
\vspace{9.0cm}
\includegraphics{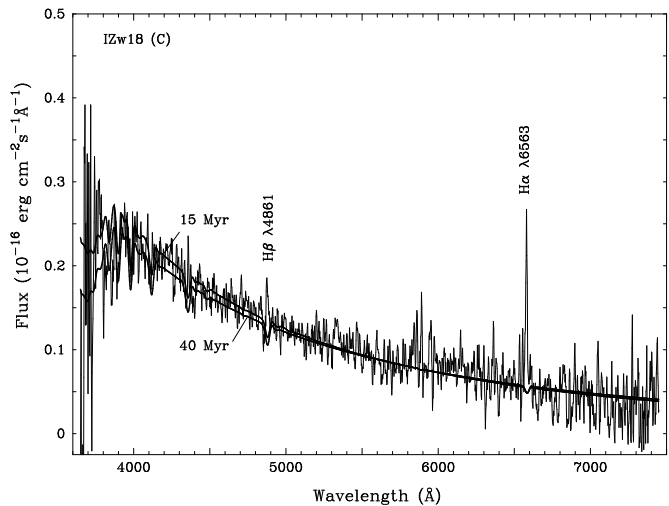}
\vspace{-0.7cm}
\caption[]{\label{Fig14}
Spectrum of the C component of I Zw 18 (thin line) on which are
superposed model stellar population SEDs with ages 15 Myr and 40 Myr (thick lines). 
The spectrum is uncorrected for extinction.}
\end{figure}

Fig. \ref{Fig14} shows the 
spectrum of the southeastern part of the C component. It is fitted very well by 
a single stellar population with age $\sim$ 15 Myr. This is in excellent 
agreement with the age derived earlier from the main-sequence turn-off in 
the CMD of the C component, supporting our large adopted distance 
of 20 Mpc to I Zw 18. To illustrate the sensitivity of synthesized SEDs to
age determination, we have also shown in Fig. \ref{Fig14} the SED of
a stellar population with age 40 Myr, the value adopted
by Dufour et al. (1996b). The agreement is not so good, the synthesized fluxes 
being systematically smaller than the observed fluxes for $\lambda$ $\la$ 
5000 \AA.

We conclude that evolutionary synthesis models further constrain the 
age of the C component to the range 15 -- 40 Myr.

%
%
\begin{figure}[t]
\vspace{9.0cm}
\includegraphics{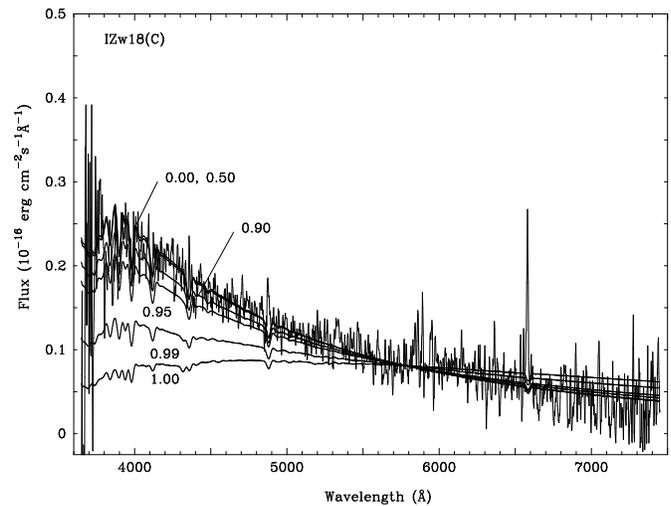}
\vspace{-0.7cm}
\caption[]{\label{Fig15}
Spectrum of the C component of I\ Zw\ 18 on which are superposed  
a modeled SED for a composite stellar population including a young stellar 
population with age 15 Myr and an older one with age 10 Gyr. Each SED is labeled
by the relative mass fraction of the old stellar population. No interstellar 
extinction correction has been applied.}
\end{figure}
%

\subsection{A 10 Gyr old stellar population?}

Although there is compelling evidence that the stellar emission 
in I\ Zw\ 18 ( in both the main body and the C component ) is due to 
stellar populations with age $\la 100$ Myr, the presence of a very faint older
stellar population with age $\sim 10$ Gyr cannot be
definitely ruled out, since such a population would not be visible
in CMDs with $V$ $\la$ 27.5 mag and would cause a non-detectable effect
in photometric and spectroscopic data.

 Figure \ref{Fig15} shows composite SEDs resulting from a mixture of a
young 15 Myr and an old 10 Gyr stellar population. Each SED is labeled by the 
mass fraction of the old stellar population. The SED with comparable masses of 
young and old stellar populations ( labeled as 0.5 ) is indistinguishable
from that of a pure young stellar population ( labeled 0 ).
However, SEDs of composite populations where the mass of the 
old stellar population is 10 times greater than
the mass of the younger stellar component can be excluded
as they are systematically redder than the observed spectrum
for $\lambda$ $\la$ 5500 \AA.

We cannot exclude therefore an underlying very old stellar population with mass 
comparable to that of the young stellar population in the C component. 
It must be however spatially coincident with the young stellar population 
because the colours are nearly constant. This is very unlikely as photometric 
studies of other BCDs (e.g. Papaderos et al. 1996ab) show redder colours 
towards the outer parts implying that the old stellar population is spatially 
distinct and more extended than the young stellar population. 

In summary, there is no need for a stellar population older than $\sim 100$ Myr 
to account for photometric and spectrophotometric properties of 
I Zw 18. Deep near - infrared 
photometric observations are needed to put stronger constraints on a older
stellar population.

\section{Age constraints from the ionized gas shell structures}

Many shell structures are seen in H$\alpha$ images of I Zw 18 
(Fig. \ref{Fig1}),
produced by the combined effect of stellar winds from massive stars and 
supernovae. Knowledge of the radii, the number of massive stars, the ambient 
density, the geometry of the gas distribution and the properties of stellar 
winds from low-metallicity massive stars and of supernovae allows,
in principle, to estimate 
the age of the stellar populations responsible for these large-scale structures.
However, this knowledge is still rather uncertain and stellar population 
age estimates from shell structures should be considered to be
more qualitative than quantitative. 

If the largest shells were produced by the evolution of the NW and SE 
components,
then the shell structures would have radii $\la$ 1200 pc. However, the centers 
of symmetry for the large shells are not coincident with the NW component.
It is more likely that the large shells are produced by older stellar clusters
as suggested by Dufour et al. (1996b), and in that case, the structures would
have radii of $\sim$ 400 -- 800 pc.

    The radius of a superbubble produced by a population of stars with
stellar winds and by supernovae is given, respectively, 
by (McCray \& Kafatos 1987):
\begin{equation}
R_S = 269 ~{\rm pc}~ (N_WL_{38}/n_0)^{1/5}t_7^{3/5}, \label{eq:wind}
\end{equation}
\begin{equation}
R_S = 97 ~{\rm pc}~ (N_{SN}E_{51}/n_0)^{1/5}t_7^{3/5}, \label{eq:sn}
\end{equation}
where $R_S$ is in pc, $N_W$ is the number of stars with stellar wind,
$L_{38}$ = $L_W$/(10$^{38}$ ergs s$^{-1}$), $L_W$ being
the mechanical luminosity of a single stellar wind, $t_7$ = $t$/(10$^7$yr),
$n_0$ is the density of the ambient gas, $N_{SN}$ is the number of supernovae,
$E_{51}$ = $E_{SN}$/(10$^{51}$ ergs), $E_{SN}$ being the mechanical energy 
of a single supernovae.

    Consider first the stellar wind mechanism. The mechanical energy of the
stellar wind of a single star in our Galaxy is $L_{38}$ $\sim$ 1 (Abbott,
Bieging \& Churchwell 1981). Taking into account the decrease of wind 
efficiency
with metallicity as $Z^{0.5}$ (Maeder \& Meynet 1994), we derive $L_{38}$ = 0.14
for I Zw 18's metallicity ( note that Hunter \& Thronson (1995) used three times 
as large a value ). Guseva, Izotov \& Thuan (1999) derived a number of 66 WR 
stars in I Zw 18. This number 
should be considered as a lower value of stars with stellar winds, 
since massive O stars can also contribute. The number density of the ambient
neutral gas $n_0$ can be estimated from H I column densities (van Zee et al. 
1998). It varies from $\sim$ 1 cm$^{-3}$ in the central part of the main body 
where the inner shell is observed, to $\sim$ 0.05 cm$^{-3}$ at the distance of 
$\sim$ 1 kpc from the main body where the largest shells are seen. 
We adopt two limiting values $n_0$ = 0.1 and 1 cm$^{-3}$, within the range of 
number densities discussed by Martin (1996). The WR stage in an instantaneous 
burst at $Z$ $\sim$ $Z_\odot$/50 is short, $t_7$ $\sim$ 0.1 (de Mello et al. 
1999). Then, the radius of the shell produced by stellar winds of WR stars is
$R_S$ $\sim$ 110 pc, if $n_0$ = 1 cm$^{-3}$ and $\sim$ 170 pc,
if $n_0$ = 0.1 cm$^{-3}$, in good agreement with the observed radius 
$\sim$ 100 pc of the 
inner shell. However, stellar winds alone cannot explain the presence of larger 
structures.

 Consider next the combined effect of supernovae. In the NW component, the 
formation of structures with radii as large as 1200 pc should be accounted for. 
For simplicity, we assume the number of supernovae to be equal to the number of 
O stars. The latter can be estimated from the total observed flux of the 
H$\beta$ emission line. The H$\beta$ flux in Table \ref{Tab3} cannot be used as 
it has not been corrected for aperture effects. Instead, we adopt the
aperture-corrected H$\beta$ flux from Guseva et al. (1999), giving a number of O 
stars equal to 4800. Adopting $t_7$ = 0.5 for the age of the main body, 
Eq. (\ref{eq:sn}) gives $R_S$ = 550 pc and 350 pc for $n_0$ = 0.1 cm$^{-3}$ and 
1 cm$^{-3}$ respectively. 

To explain the presence of structures with radii as large as 1200 pc, the age 
for the NW component should be as high as 18 -- 39 Myr, larger than the age
of $\sim$ 5 Myr inferred from ionization constraints and
the presence of WR stars. We conclude therefore that the stellar clusters in 
the NW and SE components ionizing the gas in the main body cannot be
responsible for the largest structures, and these are due to the action of 
older clusters and stellar associations. The largest shell to the west of the
NW component with radius 800 pc is likely to be connected with the stellar
association located at 4\arcsec\ to the NW of the NW component,
near its center of symmetry. If we assume that a starburst occurred
10 Myr ago, that there were as many O stars as in the NW component and
$n_0$ = 0.1 cm$^{-3}$, then $R_S$ = 840 pc. Because of the weak dependence
of $R_S$ on $E_{51}$ and $t_7$ ( Eq. \ref{eq:sn} ) even a burst 
10 times weaker with an age of 20 Myr can account for the radius of the largest
shell. Hence, star formation in the main body of I Zw 18 is likely 
to have started $\sim$ 20 Myr ago, propagating generally from the NW to the 
SE direction, and continuing the star formation started earlier in the C 
component. This age estimate is in agreement with the one derived by 
Martin (1996) with the use of a more complex model of the gas distribution,
and of the smaller distance of 10 Mpc.

\section{Conclusions} 

Our main goal here is to use observed properties of the blue compact
dwarf galaxy I Zw 18 to put constraints on its age: the high ionization state 
of the gas and the presence of WR stars in the main body, the existence 
of ionized gas in the C component, and the colour-magnitude diagrams from
HST images. We were motivated by the study of Izotov \& Thuan (1999) who have
analyzed the C/O and N/O abundance ratios of a sample of the most 
metal-deficient BCDs known, including I Zw 18. Those authors found that these
ratios are constant for galaxies with $Z$ $\la$ $Z_\odot$/20 with a very small
dispersion around the mean. This strongly suggests that intermediate-mass
stars ( $M$ $\la$ 8 $M_\odot$ ) have not had time to release their carbon and
primary nitrogen production, establishing an age upper limit of $\sim$ 100 Myr
for very metal-deficient BCDs.

The conclusion that galaxies with $Z$ $\la$ $Z_\odot$ / 20 are younger than
100 Myr has been supported by photometric and spectroscopic studies of two very
metal-deficient BCDs, SBS 0335--052 ( $Z_\odot$/40, Thuan et al. 1997,
Papaderos et al. 1998) and SBS 1415+437 ( Thuan et al. 1999a). Here
we examine the age evidence for I Zw 18. To put constraints on the age of I Zw 
18, we have followed 3 independent lines of investigation: colour-magnitude 
diagram, spectral synthesis and hydrodynamical age constraints. 
We have arrived at the following main conclusions:

1. The distance to I Zw 18 must be increased by a factor of 2 from the
previously adopted value of 10 Mpc to 20 Mpc. Such a distance is required to
have stars bright and massive enough in I Zw 18 to account for its high
state of ionization and the presence of Wolf-Rayet stars in its NW component.

2.  ($B-V$) vs. $V$ and ($V-R$) vs. $R$ CMD studies with the new distance of 20 
Mpc give ages 
derived from the main sequence turn-off of I Zw 18 of $\sim$ 15 Myr
and $\sim$ 5 Myr for the C component and main body respectively. The location
of the resolved luminous red stars with $M_R$ $\sim$ --6 mag and ($V-R$) $\sim$
0.6 -- 1.0 mag is consistent with an age $\la$ 100 Myr for the C component. The
star formation in this component is likely to have stopped 15 -- 20 Myr ago.
As for the main body, CMD analysis implies that star formation started 20 -- 50
Myr ago and still continues nowadays. Analysis of shell structures seen in
H$\alpha$ images also suggests that star formation in the main body began
$\sim$ 20 Myr ago in different locations at the NW side and has been propagating
mainly in the SE direction. The age upper limit of $\sim$ 50 Myr derived 
for the main body is a whole order of magnitude smaller that the one derived 
by Aloisi et al.
(1999) from CMD analysis of similar HST data. The difference is mainly due
to the increase in distance of I Zw 18 by a factor of 2.

3. Fits to the spectral energy
distributions give ages of $\sim$ 5 Myr for the main body and $\sim$ 15 -- 40
Myr for the C component.

In summary, all three lines of investigation ( CMDs, the distribution
of H$\alpha$ shells and 
spectroscopy ) lead to the same conclusion, that I Zw 18 did not start to 
form stars until $\la$ 100 Myr ago. This supports the contention of Izotov \&
Thuan (1999) that all very metal-deficient galaxies ( $Z$ $\la$ $Z_\odot$/20 )
are young.

\begin{acknowledgements} 
Y.I.I. and N.G.G. thank the Universit\"ats--Sternwarte of G\"ottingen 
and Y.I.I. thanks the University of Virginia for warm hospitality.
We are grateful to D. Schaerer for sending to us his stellar 
evolutionary synthesis models in electronic form and A. Aloisi for 
communicating her results.
We acknowledge the financial support of Volkswagen Foundation Grant No. I/72919
(Y.I.I., N.G.G., P.P. and K.J.F.) and of National Science Foundation grants
AST-9616863 ( T.X.T. and Y.I.I. ) and AST-9803072 ( C.B.F. ).
Research by K.J.F. and P.P. has been supported by Deutsche Agentur 
f\"{u}r Raumfahrtangelegenheiten (DARA) GmbH grants 50\ OR\ 9407\ 6
and 50\ OR\ 9907\ 7. 
\end{acknowledgements}

{}

\end{document}